\documentstyle[twocolumn,aps,amsmath,amssymb,epsf,times]{revtex}

\def\beq{\begin{equation}}
\def\eeq{\end{equation}}
\def\bea{\begin{eqnarray}}
\def\eea{\end{eqnarray}} 

\def\ga{\gamma} 
 \def\De{\Delta}
  \def\eps{\epsilon}

\def\pa{\partial}

\def\half{\mbox{$\frac{1}{2}$}}

\begin{document}
\preprint{\framebox{NIKHEF-99-017, KVI-1466}}
\draft

\title{Field theory of nucleon to higher-spin
       baryon transitions}

\author{Vladimir Pascalutsa$^{a,c}$ and Rob Timmermans$^{b,c}$}
\address{$^a$Theory Group, NIKHEF, University of Amsterdam,
         P.O. Box 41882, 1009 DB Amsterdam, The Netherlands}
\address{$^b$Theory Group, KVI, University of Groningen,
         Zernikelaan 25, 9747 AA Groningen, The Netherlands}
\address{$^c$Institute for Theoretical Physics, 
         University of Utrecht, Princetonplein 5,
         3584 CC Utrecht, The Netherlands}

\date{\today}
\maketitle

\begin{abstract}
We discuss the nucleon to higher-spin $N$- and $\Delta$-resonance
transitions by pions and photons. The higher-spin baryons are
described by Rarita-Schwinger fields and, as we argue, this
imposes a stringent consistency requirement on the form of
the couplings. Popular $\pi N\Delta$ and $\gamma N\Delta$
couplings are inconsistent from this point of view. We construct
examples of consistent interactions with the same nonrelativistic
limit as the conventional ones.
\end{abstract}

Current experimental efforts promise to greatly advance the
understanding of the strong and electroweak structure and the
in-medium properties of the nucleon and its excited $N^*$-states,
such as the spin-3/2 $\Delta$-isobar~\cite{Ber96,Han98,Wei98}. In
theoretical studies, an important role is played by relativistic
Lagrangians describing on a hadronic level the interactions of
the baryons with pions and photons. They are often used, at tree
level or in a unitarized model, to extract
the relevant coupling constants from the data, which can then be
compared to results obtained in quark or other ``QCD-inspired''
microscopic models. To do this in a meaningful manner, one needs
consistent interactions between nucleons, $N^*$-resonances,
and pions and photons.

However, modern treatments of higher-spin ($s\geq3/2$) baryon fields
within the standard Rarita-Schwinger (RS) formalism are problematic.
The difficulties are generic to any field-theoretic description
of higher-spin particles and are related to the fact that (in
a relativistic, local formulation in four space-time dimensions)
a higher-spin field contains more components than is needed to
represent the spin degrees of freedom (DOF) of the particle. 
The standard free-field formulations~\cite{Rar41,Sin74,Fro78} 
are given by Lagrangians which, in addition to the Dirac- or
Proca-type equations, yield constraint equations that reduce the
number of independent components of the field to the correct value.
The issue is how to introduce interactions: When these
are not constructed consistently with the free theory, the
constraints may be violated and consequently 
the unphysical extra DOF will become involved.
The widely-used $\pi N\Delta$ and $\gamma N\Delta$ interactions
given below in Eqs.~(\ref{eq:conv1}) and (\ref{eq:conv2}) are
examples of such inconsistent couplings. The pathologies of the
$\pi N\Delta$ coupling have been especially thoroughly
discussed~\cite{Nat71,Hag71,Sin73,Ben89,Pas98}. It is
the purpose of this Rapid Communication 
to present a remedy
for these problems, construct explicit examples of alternative
consistent interactions, and illustrate some consequences.

An elegant general way to distinguish consistent theories for
high-spin fields is to use the correspondence between the
local symmetries and the DOF content of the theory (see
the theorem quoted below). The free massless theory can be
constructed by demanding the action to be invariant under a
number of gauge transformations~\cite{Cur79,ArD80,WiF80},
constraining the number of DOF to two. The mass term breaks these symmetries
such that the number of DOF is raised to the appropriate $2s+1$.
Our basic premise is that a consistent interaction should not
``activate'' the spurious DOF, and therefore the full interacting
theory must obey similar symmetry requirements as the corresponding
free theory.

It is sometimes possible to formulate interactions which destroy
the symmetries of the massless free theory, but in a way similar to
the mass term.  For instance, the ``minimal'' electromagnetic 
coupling of the RS field, or the conventional $\pi N\Delta$ coupling
of Eq.~(\ref{eq:conv1}) with the specific choice $z_\pi=1/2$, are
interactions of this type. It then appears that the constraints can
be violated only for specific values of the interaction strength.
However, such theories in general have non-positive-definite
commutators -- the Johnson-Sudarshan problem~\cite{Joh61,Hag71}, as well
as acausal propagations -- the Velo-Zwanziger problem~\cite{Vel69,Sin73}.
Also, correct derivations of Feynman rules in these theories 
indicate that Lorentz invariance is not obvious \cite{Yam86,Pas98} 
despite the fact that one starts from
a manifestly Lorentz-invariant Lagrangian.
We shall not further discuss these problems in here, but
merely assume that they should be absent in a consistent theory.
We therefore adopt the viewpoint
that consistent interactions must support the local symmetries
of the free massless RS formulation, while
the mass term breaks these symmetries in the correct manner.

In first instance, the interactions can be chosen to simply
preserve the gauge symmetries of the free massless theory. In fact,
the possibility to construct consistent higher-spin field theories
with such gauge-invariant (GI) couplings was pointed out by
Weinberg and Witten already some time ago~\cite{Wei80}, but
apparently has never been exploited in hadronic physics.
Here we shall confine ourselves to exploring this road,
which is sufficient for formulating consistent
$N\stackrel{\pi,\gamma}{\rightarrow}N^\ast$ interactions.

To present our arguments more systematically, let us first recall
that within the RS formalism a field of spin $s=j+1/2$ ($j$ is
an integer) is represented by a symmetric Lorentz tensor-spinor
$\psi_{\mu_1\ldots\mu_j}^{(\alpha)}(x)$ of rank $j$; the spinor
index $\alpha$ will be omitted in what follows. Note that such a 
field has
\begin{equation}
   C_j \equiv 4(j+1)(j+2)(j+3)/6
\end{equation}
independent components. The requirement that the field describes a massless
particle, with only two helicities, leads to an essentially unique
definition of the theory~\cite{Fro78,Cur79,ArD80,WiF80}. Namely,
the action must be invariant under the gauge transformation
\begin{equation}
\label{eq:gauge}
   \delta\psi_{\mu_1\ldots\mu_j} = 
     (1/j)\left[\partial_{\mu_1}\epsilon_{\mu_2\ldots\mu_j}+\ldots+
     \partial_{\mu_j}\epsilon_{\mu_1\ldots\mu_{j-1}}\right] \ ,
\end{equation}
where $\epsilon(x)$ is a symmetric tensor-spinor field of
rank $j-1$, subject to the traceless condition: 
$\gamma^{\mu_1}\epsilon_{\mu_1\mu_2\ldots\mu_{j-1}}=0$.
Furthermore, the field itself must satisfy the
gauge-invariant condition
\begin{equation}
\label{eq:trless}
  g^{\mu_1\mu_2}\gamma^{\mu_3}\psi_{\mu_1\mu_2\mu_3\ldots\mu_j} = 0 \ .
\end{equation}

To count the number of DOF we may use the Hamiltonian
framework~\cite{Dir64,Git86,Hen92,Wei95}, where, given the
Lagrangian density for $\psi$, one defines the conjugate momentum
  $\pi^{\mu_1\ldots\mu_j}=
  \partial{\mathcal L}/\partial\dot{\psi}_{\mu_1\ldots\mu_j}$
and determines all the constraints
in the ($\pi,\psi$) phase space of the theory. Taking into
account the condition Eq.~(\ref{eq:trless}), the field actually has
\begin{equation}
   N_{\rm comp} = C_{j}-C_{j-3} = 6j(j+1)+4
\end{equation}
independent components, and so does its conjugate momentum.
However, only two (for each of them) are needed to
describe a massless particle. The rest is
to be eliminated by means of the phase-space constraints.
Let $N_I$ and $N_{I\!I}$ denote the number of first- and second-class
constraints, respectively. Each first-class (second-class)
constraint eliminates two (one) DOF~\cite{Hen92}. Thus,
in a theory with only physical DOF $N_I$ and $N_{I\!I}$ must
satisfy
\begin{equation}
\label{eq:count}
    2N_{\rm comp} - 2N_I - N_{I\!I} = 2 + 2  \ .
\end{equation}
An explicit determination of the constraints in the most general
case by, for instance, the usual Dirac-Bergmann procedure \cite{Dir64}
is a formidable task. But their number can easily be assessed
by using the following theorem, which 
establishes a precise correspondence between the local 
symmetries of the action and the first-class constraints:

\noindent
{\em Theorem}.
A Lagrangian theory invariant under a local transformation
with $n$ independent parameters has $n$ primary 
first-class constraints~\cite{Git86}; and the total
number of first-class constraints is $(d+1)\times n$, where $d$
is the highest order of the time-derivative operator acting
on the parameters of the transformation~\cite{Vas99}.

In our case, the Lagrangian is invariant under Eq.~(\ref{eq:gauge}),
hence $d=1$ and $n=C_{j-1}-C_{j-2}=2j(j+1)$, while
\begin{equation}
\label{eq:first}
   N_I = 4j(j+1)
\end{equation}
by the second part of the theorem. Furthermore, since fermionic
theories are of first order in space-time derivatives, the total
number of primary constraints is equal to the number of field
components, $N^{(1)}_I+N^{(1)}_{I\!I}=N_{\rm comp}$.
Invoking the first part of the theorem we have
$N^{(1)}_I=n$. From the fact that (in the massless case)
there are only primary second-class constraints, {\em i.e.},
$N_{I\!I}=N^{(1)}_{I\!I}$, we find 
\begin{equation}
\label{eq:second}
   N_{I\!I} = N_{\rm comp}-n = 4j(j+1)+4 \ .
\end{equation}
By using these values for $N_I$ and $N_{I\!I}$, one can
check that Eq.~(\ref{eq:count}) is indeed satisfied.
We have thus proven the unitarity, or the so-called
``no-ghost'' theorem~\cite{Fro78,Cur79}, of the 
massless higher-spin fermion formulation. A similar 
proof applies to the formulation for higher-spin bosons.

The mass term is usually introduced so as to break the gauge 
symmetry, turning all the first-class constraints into the
second class. The resulting number of the second-class
constraints $\bar{N}_{I\!I}$ must provide the physical 
DOF counting
\begin{equation}
\label{eq:mcount}
    2N_{\rm comp} - \bar{N}_{I\!I} = 2(2s+1)  \ .
\end{equation} 
One can in general write $\bar{N}_{I\!I}=N_I+N_{I\!I}+N'_{I\!I}$,
and find from Eqs.~(\ref{eq:first})--(\ref{eq:mcount}) that
$N'_{I\!I}=4j^2$.
This shows that the mass term should play a rather subtle role:
besides turning the first-class constraints of the massless
theory into the second class, some number $N'_{I\!I}$ of
new second-class constraints must be generated. 

The couplings consistent with the above free-theory construction,
for both massless and massive cases, will apparently be only those
which are invariant under transformation Eq.~(\ref{eq:gauge}) or
its modifications, the so-called ``deformations,'' with the same
$d$ and $n$. Realizations
based on deformations of the free-theory symmetries appear
to be unavoidable in the construction of ``minimal'' couplings
of the RS fields to the photon or gravity. However, it
is not necessary for constructing consistent $N\rightarrow N^\ast$
transition interactions. In this case, one can generally construct
couplings invariant under Eq.~(\ref{eq:gauge}).

To illustrate all this, let us specify the discussion to the
spin-3/2 case, relevant to the important example of the
$\Delta$-isobar.
The spin-3/2 field is described by the sixteen-component
vector-spinor $\psi^\mu(x)$,
with for the massless case the Lagrangian density
\begin{equation}
\label{eq:free}
  {\mathcal L} = \bar{\psi}^{\lambda}
  {\cal O}_{\lambda\varrho}(a)\,\mbox{\small{$\frac{1}{2}$}}
  \left\{\sigma^{\varrho\mu},
  i\partial\hspace{-2mm}\slash\right\}\,
  {\cal O}_{\mu\nu}(a)\psi^\nu \ ,
\end{equation}
where 
\begin{equation}
  {\cal O}_{\mu\nu}(a) \equiv \exp\left(
  \mbox{\small{$\frac{1}{4}$}}a\gamma_\mu\gamma_\nu\right) =
  g_{\mu\nu}+\mbox{\small{$\frac{1}{4}$}}(e^a-1)\gamma_\mu\gamma_\nu  \ ,
\end{equation} 
and the arbitrary constant $a$ represents the freedom due to
the point-transformation invariance~\cite{Nat71,Ben89};
$\sigma^{\varrho\mu}=[\gamma^\varrho,\gamma^\mu]/2$. The action
of this theory is invariant under the gauge transformation
\begin{equation}
\label{eq:gauget}
  \delta\psi_\mu = {\cal O}_{\mu\nu}(-a)\,
                    \partial^\nu\eps \ ,
\end{equation}
where $\eps(x)$ is a spinor field. By using the theorem
we have $N_I=8$ and $N_{I\!I}=12$, which is of course in
agreement with an explicit evaluation of the constraints 
\cite{Pas98,Sen77}. 

For $|a|<\infty$, the tensors ${\cal O}(a)$ form a group with
${\cal O}_{\mu\nu}(0)=g_{\mu\nu}$ as the unit element and product
law ${\cal O}_\mu^\varrho(a_1){\cal O}_{\varrho\nu}(a_2)
     ={\cal O}_{\mu\nu}(a_1+a_2)$. 
Different choices among these finite values of $a$
amount to the field redefinition
    $\psi_\mu'= {\cal O}_{\mu\nu}(a)\,\psi^\nu$.
Since $\det{\cal O}(a)=e^a$ is a constant, any choice can be
made without affecting the $S$ matrix. 

For the ``forbidden'' value $a=-\infty$, the Lagrangian becomes
\begin{equation}
\label{eq:gamu}
  {\mathcal L} =
       i\bar{\psi}^\lambda{\cal O}_{\lambda\mu}(-\infty)
       \gamma_\varrho {\cal O}^{\mu\nu}(-\infty)
       \partial^\varrho\psi_\nu \ ,
\end{equation}
where ${\cal O}_{\mu\nu}(-\infty)=g_{\mu\nu}
       -\mbox{\small{$\frac{1}{4}$}}\gamma_\mu\gamma_\nu$.
This is the massless version of the theory recently considered
by Haberzettl~\cite{Hab98}. Determining the constraints
for this case we find $N_I=4$ and
$N_{I\!I}=12$, hence more DOF than the RS theory. This can
be understood by observing that the Lagrangian Eq.~(\ref{eq:gamu})
is invariant under $\delta\psi_\mu=\gamma_\mu\eps$,
{\it i.e.}, it has the same number of local transformations
as the RS theory, but without the space-time derivative.
Note that the massive case of this $a=-\infty$ theory~\cite{Hab98} 
has the same $\delta\psi_\mu=\gamma_\mu\eps$ symmetry.

The massive RS theory is obtained by the replacement
$   \partial^\mu \rightarrow \partial^\mu
        + \mbox{\small{$\frac{1}{4}$}} iM\,\gamma^\mu $
in Eq.~(\ref{eq:free}). The mass term breaks the gauge
symmetry Eq.~(\ref{eq:gauget}) in the correct way, raising
the number of DOF to four as is appropriate for a massive 
spin-3/2 particle.
The propagator of the theory is the well-known RS propagator;
in terms of spin-projection operators $P^{(J)}$~\cite{Fro78,Nie81}
it reads\footnote{In what follows, we choose $a=0$ without loss
of generality.}
\begin{eqnarray}
\label{eq:RSprop}
   S_{\mu\nu}(p) & = & \frac{1}{p\hspace{-1.65mm}\slash-M+i\varepsilon}
     P_{\mu\nu}^{(3/2)}-\frac{2}{3M^2}(p\hspace{-1.65mm}\slash+M)
     P^{(1/2)}_{22,\mu\nu} \nonumber \\
     & & + \,\frac{1}{\sqrt{3}M} \left(P^{(1/2)}_{12,\mu\nu}
         + P^{(1/2)}_{21,\mu\nu}\right) \ ,
\end{eqnarray}
where
\begin{equation}
   P^{(3/2)}_{\mu\nu} = g_{\mu\nu} - \frac{1}{3}\gamma_\mu\gamma_\nu
     - \frac{1}{3p^2} (p\hspace{-1.65mm}\slash\gamma_\mu p_\nu
        + p_\mu\gamma_\nu p\hspace{-1.65mm}\slash)
\end{equation}
projects onto the pure spin-3/2 states, while
\begin{eqnarray}
\label{eq:shalf}
    {P}^{(1/2)}_{22,\mu\nu} & = &
                 p_\mu p_\nu/p^2        \ , \nonumber \\
    {P}^{(1/2)}_{12,\mu\nu} & = & p^\varrho p_\nu
                 \sigma_{\mu\varrho}/(\sqrt{3}\,p^2) \ , \\ \nonumber
    {P}^{(1/2)}_{21,\mu\nu} & = & p_\mu p^\varrho
                 \sigma_{\varrho\nu}/(\sqrt{3}\,p^2) \ ,
\end{eqnarray}
are projection operators onto the spin-1/2 sector of the RS
theory. The pole part of the RS propagator is proportional
to $P^{(3/2)}$, while the nonpole part involves the spin-1/2
sector.

Consider next the interactions. In the literature, a popular
choice for the $\pi N\Delta$ coupling is\footnote{For brevity,
isospin space is omitted throughout.}
\begin{equation}
\label{eq:conv1}
   {\mathcal L}_{\pi N\Delta} =
   (f_{\pi N\Delta}/m_\pi)\,\bar{\psi}^\mu \Theta_{\mu\nu}
   (z_\pi)\Psi\,\partial^\nu\phi + {\rm H.c.} \ ,
\end{equation}
while for the $\gamma N\Delta$ couplings one often takes
\begin{eqnarray}
\label{eq:conv2}
 {\mathcal L}^{(1)}_{\gamma N\Delta} & = &
        \frac{ieG_1}{2m} \bar{\psi}^\varrho
        \Theta_{\varrho\mu}(z_{\gamma,1})\,
        \gamma_{\nu}\gamma_5\Psi F^{\mu\nu} + \mbox{H.c.} \ ,
        \nonumber \\
 {\mathcal L}^{(2)}_{\gamma N\Delta} & = &
        -\frac{eG_2}{(2m)^2} \bar{\psi}^\varrho
        \Theta_{\varrho\mu}(z_{\gamma,2})\,
        \gamma_5\,\partial_\nu\Psi\,F^{\mu\nu} + \mbox{H.c.} \ ,
         \\ 
 {\mathcal L}^{(3)}_{\gamma N\Delta} & = &
        -\frac{eG_3}{(2m)^2}\bar{\psi}^\varrho
        \Theta_{\varrho\mu}(z_{\gamma,3})\,
        \gamma_5\Psi\,\partial_\nu F^{\mu\nu} + \mbox{H.c.} \ ;
        \nonumber
\end{eqnarray}
see, {\it e.g.}, Refs.~\cite{Pec68,Hoh83,Ols74}.
Here, $\psi^\mu$, $\Psi$, and $\phi$ denote the $\Delta$-isobar
vector-spinor, nucleon spinor, and pion pseudoscalar fields,
with masses $M$, $m$, and $m_\pi$, respectively; $F^{\mu\nu}$
is the photon field tensor; $e\simeq\sqrt{4\pi/137}$ is the
proton charge; $f_{\pi N\Delta}$ and $G_i$ ($i=1,2,3$) are
dimensionless coupling constants.
For real photons, only the $G_1$ and $G_2$ terms contribute.
The interactions in Eqs.~(\ref{eq:conv1}) and (\ref{eq:conv2})
all contain the tensor
 $ \Theta_{\mu\nu}(z) = g_{\mu\nu} -
   (z+\mbox{\small{$\frac{1}{2}$}})\gamma_\mu\gamma_\nu $;
the constants $z_\pi$ and $z_{\gamma,i}$ ($i=1,2,3$) with
arbitrary values are the so-called ``off-shell parameters.''

These ``conventional'' $\pi N\Delta$ and $\gamma N\Delta$
interactions are inconsistent with the free spin-3/2 RS theory,
for any value of the off-shell parameters, see, {\it e.g.},
Refs.~\cite{Nat71,Hag71,Sin73,Pas98}. They do
not possess any local symmetries of the RS field, and as
a consequence they violate the constraints and involve the
unphysical lower-spin DOF. The latter contribute to the
observables in terms of the ``spin-1/2 backgrounds''~\cite{Ols74}.

\begin{figure}[t]
\begin{center}
\epsffile{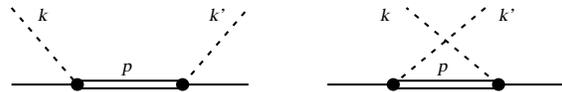}
\end{center}
\caption{Typical $\Delta$-exchange graphs. Here the dashed, solid,
    and double lines denote the pion, the nucleon, and the $\Delta$
    propagations, respectively.}
\label{fig:1}
\end{figure}

In contrast, $N\stackrel{\pi,\gamma}{\rightarrow}\Delta$
couplings which are invariant under the gauge transformation
Eq.~(\ref{eq:gauget}) will be fully consistent in that sense.
Such GI couplings can easily be
constructed by using the manifestly invariant RS field tensor
\begin{equation}
  G^{\mu\nu}=\partial^\mu\psi^\nu -\partial^\nu\psi^\mu
\end{equation}
and its dual
 $ \tilde{G}^{\mu\nu} = \mbox{\small{$\frac{1}{2}$}}
         \varepsilon^{\mu\nu\varrho\sigma}G_{\varrho\sigma}$.
The corresponding vertices $\Gamma^\mu(p,p-k,k)$,
where $p$ and $k$ are the momenta of the $\Delta$ and,
{\it e.g.}, the pion, while $\mu$ is the Lorentz index
associated with the $\Delta$-field, will satisfy
\begin{equation}
   p_\mu \Gamma^\mu(p,p-k,k) = 0 \ .
\end{equation}
  From Eqs.~(\ref{eq:RSprop}) and (\ref{eq:shalf}) one
can then immediately see that all nonvanishing $\Delta$-exchange 
amplitudes (see, {\it e.g.}, Fig.~\ref{fig:1}),
$$
   \Gamma^\mu(p,p-k',k')\,
      S_{\mu\nu}(p)\,\Gamma^\nu (p,p-k,k) \ ,
$$
are proportional to the spin-3/2 projection operator,
and thus the unphysical spin-1/2 sector decouples.
(Let us remark here that it was correctly
anticipated in Refs.~\cite{Wil85,Ade86}
that $\Delta$-exchange amplitudes should be
proportional to the spin-3/2 projection operator. However, the
{\em ad-hoc} prescriptions used in these works cannot be derived
from a local Lagrangian and face other problems~\cite{Ben89}.
They are also at variance with the standard result~\cite{Gre87}
that the amplitude for exchange of a spin-$s$ particle must behave
as $p^{2s}/(p^2-m^2)$. These criticisms do not apply to the present
approach, where a typical $\Delta$-exchange amplitude is given by
Eq.~(\ref{eq:ampl}) below.)

In principle, there are many GI couplings one can construct.
We will focus here only on the ones that become equivalent to
the conventional couplings at the $\Delta$-pole and hence have
the same nonrelativistic limit. By using this ``pole equivalence''
we can establish the connection with coupling constants used
in the vast number of previous studies, including the 
nonrelativistic~\cite{KMO84} and heavy-baryon~\cite{JeM91,HHK97} formalisms.
 
For the $\pi N\Delta$ interaction we take
\begin{equation}
\label{eq:int3}
  {\mathcal L}_{\pi N\Delta} =
     f\,\bar{\Psi}\gamma_5\gamma_\mu
     \tilde{G}^{\mu\nu}\partial_\nu\phi + \mbox{H.c.}
\end{equation}
The pole equivalence implies that the vertex $\Gamma^\mu$
obtained from this Lagrangian, when contracted with the free
RS vector-spinor $u_\mu(\hat{p})$, where $\hat{p}$ is the
on-shell momentum of the $\Delta$, $\hat{p}^2=M^2$,
becomes equivalent to the
conventional $\pi N\Delta$ vertex, $\Gamma^\mu_{\rm conv}$,
found from Eq.~(\ref{eq:conv1}), {\it i.e.},
\begin{equation}
   \Gamma^\mu(\hat{p},\hat{p}-k,k)\, u_\mu(\hat{p}) =
   \Gamma^\mu_{\rm conv}(\hat{p},\hat{p}-k,k) \,
   u_\mu(\hat{p}) \ .
\end{equation}
This condition requires us to identify the coupling constant
in Eq.~(\ref{eq:int3}) as $f=f_{\pi N\Delta}/(m_\pi M)$, in
terms of the coupling $f_{\pi N\Delta}$ of Eq.~(\ref{eq:conv1}).
However, we emphasize that despite the imposed pole
equivalence, the two couplings will give different results
for $\Delta$-exchange amplitudes of, {\it e.g.}, Fig.~\ref{fig:1},
even at the pole $p^2=M^2$. This is because with
the conventional coupling one still encounters the background
due to the negative-energy state contribution of
the spin-1/2 sector.
In contrast, using the GI interaction Eq.~(\ref{eq:int3}),
we obtain the amplitude
\begin{eqnarray}
\label{eq:ampl}
   && \Gamma^\mu(p,p-k',k')\,S_{\mu\nu}(p) \,
      \Gamma^\nu(p,p-k,k) \nonumber \\
   && \hspace{2cm} =
      \frac{(f_{\pi N\Delta}/m_\pi)^2}{p\hspace{-1.65mm}\slash-M} \, 
      \frac{p^2}{M^2}\, P^{(3/2)}_{\mu\nu}\, {k'}^\mu k^\nu \ ,
\end{eqnarray}
for any $\Delta$-momentum $p$. 
Some realistic calculations of $\pi N$
scattering lengths and phase shifts using this amplitude 
have recently been reported~\cite{Lah99,PaT99}. These studies
indicate large qualitative differences with the conventional
approach, while in both approaches agreement with experiment
can be achieved due to the interplay of other reaction mechanisms.

Considering the photon couplings, the GI $\gamma N\Delta$
interactions that are lowest in number of derivatives read
\begin{eqnarray}
\label{eq:newgamma}
  {\mathcal L}_{\gamma N\Delta} & = & e\bar{\Psi} \left(
  g_1\tilde{G}_{\mu\nu} + g_2 \gamma_5 G_{\mu\nu} +
  g_3\gamma_\mu\gamma^\varrho\tilde{G}_{\varrho\nu}\right.
  \nonumber \\ && \left. +
  g_4\gamma_5\gamma_\mu\gamma^\varrho G_{\varrho\nu}
  \right)F^{\mu\nu} + \mbox{H.c.}
\end{eqnarray}
The first term contributes purely to the magnetic-dipole
transition in the Sachs-type decomposition
of the $\gamma N\Delta$ vertex~\cite{Jon73}. The second term is
up to a total derivative equal to the sum of the conventional
$G_2$ and $G_3$ couplings of Eq.~(\ref{eq:conv2}),
provided that $G_2=G_3=(2m)^2 g_2$
and $z_{\gamma,2}=z_{\gamma,3}=-1/2$. Therefore, for real photons
the $g_2$ coupling and the $G_2$ coupling with $z_{\gamma,2}=-1/2$
are fully equivalent. The $g_3$ and $g_4$ terms are new. However,
at the $\Delta$-pole the $g_3$ and $G_1$ couplings become equivalent,
provided that $g_3=G_1/(2mM)$. The same applies to the $g_4$ term.
Thus, the contribution of the GI couplings to the magnetic-dipole
$G_{\!M}$ and the electric-quadrupole $G_{\!E}$ transition form
factors at the $\Delta$-pole reads,\footnote{A complete treatment
of the Coulomb (longitudinal) quadrupole requires higher derivatives
than in Eq.~(\ref{eq:newgamma}).} in the conventions of
Ref.~\cite{Jon73},
\begin{eqnarray}
   3G_{\!M} &=& 2m(M+m) g_1 - m(M-m) g_2 \nonumber \\
            & & + m(3M+m) (g_3+g_4) \nonumber \ , \\
   3G_{\!E} &=& m(M-m)(g_3+g_4-g_2) \ .
\end{eqnarray}

Since the considered $\gamma N\Delta$ couplings are invariant
under both the electromagnetic and the RS gauge transformation,
the corresponding vertex $\Gamma^{\mu\varrho}(p,q)$, where $q$
is the photon momentum, obeys the transversality condition
with respect to both indices, {\it i.e.},
\begin{equation}
  p_\mu \Gamma^{\mu\varrho}(p,q) =
  q_\varrho\Gamma^{\mu\varrho}(p,q) = 0 \ .
\end{equation}
Note that we have excluded couplings that contain
$\sigma_{\mu\nu}G^{\mu\nu}$. Such couplings project
onto the purely spin-1/2 contribution ($\ga_{\mu}\ga_{\nu}$ can be
written in terms of the spin-1/2 projection operators only),
which on the other hand decouple because of the gauge symmetry.
Hence, these couplings lead to vanishing amplitudes, see 
Ref.~\cite{Pas98} for an example.

An interesting extension is to study interactions that are
not exactly gauge invariant, but the variation of which is
proportional to some free-field equations. 
The invariance of the full action can then be provided by
a variation of a corresponding free action, as illustrated below.
In such theories, the
decoupling of the unphysical DOF happens only when the
particles, the free-field equations of which become involved,
are on their mass shell. An analogous
situation arises, for instance, in QED where the electromagnetic
currents are conserved only when the external lepton legs are
on-shell. This can be physically acceptable since the spurious
DOF, even though present off-shell, do not contribute to
observables. Moreover, such interactions may have the advantage
of being lower in number of derivatives than the explicitly 
gauge-invariant ones.

However, for the $N\rightarrow\Delta$ case our attemps to find 
an interaction of this type led only to (locally) supersymmetric 
realizations. Consider, {\it e.g.},
\begin{equation}     
  {\mathcal L} = g\bar{\Psi}\gamma_\mu 
        (i\partial\hspace{-2mm}\slash\phi+m\phi)\psi^\mu \ ,
\end{equation}
where for now all the fields are Hermitian.
Under the variation $\delta\psi_\mu=\partial_\mu\eps$, we
have, up to a total derivative,
\bea
\label{eq:vary}
  \delta{\mathcal L} & = & -g\left[
  (i\pa_\mu \bar{\Psi}\gamma^\mu+m\bar{\Psi})
  (\partial\hspace{-2mm}\slash\phi-im\phi) \right. \nonumber \\
  & & + \left. i\bar{\Psi}(\partial^2\phi+m^2\phi)\right] \eps \ ,
\eea
which is indeed proportional to the free-field equation,
if the pseudoscalar and the spinor field have the same mass 
equal to $m$. This variation is cancelled by the variation of the
free Lagrangian,
$
{\mathcal L}_0 = \half \partial_\mu\phi\,\partial^\mu\phi
-\half m^2\phi+\half \bar{\Psi}(i\partial\hspace{-2mm}\slash-m)\Psi$,
under the local transformation
$$
\delta\phi = ig \bar{\Psi}\eps, \,\,\,
\delta\Psi = g [\partial\hspace{-2mm}\slash\phi-im\phi]\eps \ ,
$$
which obviously is a supersymmetric transformation.
It can still be suspected that this model is not fully consistent:
a nontrivial supersymmetry necessitates the balance between
fermionic and bosonic DOF, which may thus require
inclusion of more boson fields as well as more interaction terms, see,
{\em e.g.}, \cite{Nie81}.
However, we shall not pursue this line here. 
As far as $N\rightarrow\Delta$ couplings are concerned, a development of
such supersymmetric models seems neither necessary nor promising at present.

In conclusion, we have shown how the requirement of gauge invariance 
allows one to incorporate both manifest covariance and consistent DOF
counting in a local high-spin field formulation. We have therefore
proposed to use the gauge-invariant interactions for describing
the various meson- and photon-induced $N\rightarrow N^\ast$
transitions. This approach has been illustrated by the example
of a spin-3/2 $N^\ast$-resonance, the $\Delta$-isobar.
While the conventional interactions used to describe the nucleon
to $\Delta$-isobar transitions by pions and photons are well-known
to be inconsistent, we have constructed explicit examples of novel
consistent interactions that become equivalent to the conventional
ones at the $\Delta$-pole and in the nonrelativistic limit. We
emphasize that, even though most of our discussion has been focused
on the interactions {\it explicitly} invariant under the free-field
transformation Eq.~(\ref{eq:gauge}), one can possibly construct 
consistent interactions based on its deformations with the same
parameters. Moreover, interactions which are invariant up to 
some free-field equations can also be consistent, but in
our attempts to construct such a $\pi N\De$ coupling we were led
to a locally supersymmetric realization which has an obscure
phenomenological implementation. On the other hand, our results
concerning the explicitly gauge-invariant interactions are
certainly relevant to theoretical studies related to the ongoing 
experimental programs on meson-, photo-, and electroproduction
of $N^*$-resonances.

\acknowledgments
V.~P. would like to thank B.~de~Wit, J.~W. van Holten,
A.~Waldron for a number of illuminating discussions, 
and J.~H.~Koch for helpful remarks on the manuscript; 
R.~T. is grateful to J.~Weda and O.~Scholten for useful conversations.
The research of V.~P. was supported by the Netherlands
Research Organization (NWO) via FOM, while that of R.~T. was made
possible by a fellowship of the Royal Netherlands Academy of Arts
and Sciences.

\end{document}